\documentclass[aps,prl,reprint,showpacs]{revtex4-1}
\usepackage{graphicx}
\draft % marks overfull lines with a black rule on the right
\usepackage{bm}
\pdfoutput=1

\def\vb#1{{\bm#1}}
\def\v#1{\mathbf{#1}}			

\def\vOmega{\vb{\Omega}}

\def\vsigma{\vb{\sigma}}

\def\ls{\lambda_{\rm s}}
\def\tls{\tilde{\lambda}_{\rm s}}
\def\kt{k_{t}}
\def\ks{k_{s}}

\def\tsf{\tau_{\rm sf}}
\def\ttsf{\tilde{\tau}_{\rm sf}}

\def\grad{{\rm grad \,}}
\def\div{{\rm div \,}}

\def\del{\partial}

\begin{document}

%\preprint{}

%\title{Spin-current generation from surface acoustic wave in nonmagnets} %Title of paper
%
\title{Mechanical generation of spin current by spin-rotation coupling}



\author{Mamoru Matsuo$^{1,2}$,
Jun'ichi Ieda$^{1,2}$, 
Kazuya Harii$^{1,2}$,
Eiji Saitoh$^{1,2,3,4}$,
and Sadamichi Maekawa$^{1,2}$ }
\affiliation{
$^{1}$Advanced Science Research Center, Japan Atomic Energy Agency, Tokai 319-1195, Japan \\
$^{2}$CREST, Japan Science and Technology Agency, Sanbancho, Tokyo 102-0075, Japan\\
$^{3}$Institute for Materials Research, Tohoku University, Sendai 980-8577, Japan \\
$^{4}$WPI, Advanced Institute for Materials Research, Tohoku University, Sendai 980-8577, Japan
}

\date{\today}

\begin{abstract}
Spin-rotation coupling, which is responsible for angular momentum conversion between the electron spin and rotational deformations of elastic media, is exploited for generating spin current.
This method requires neither magnetic moments nor spin-orbit interaction. 
The spin current generated in nonmagnets is calculated in presence of surface acoustic waves. 
We solve the spin diffusion equation, extended to include spin-rotation coupling, and find that larger spin currents can be obtained in materials with longer spin lifetimes.
Spin accumulation induced on the surface is predicted to be detectable by time-resolved Kerr spectroscopy.

\end{abstract}


\maketitle %\maketitle must follow title, authors, abstract and \pacs

\paragraph{Introduction.---}%%%%%%%%%%%%%%%%%%%%%%%%%%%
Spin current, a flow of spins, is a key concept in the field of spintronics\cite{MaekawaEd2006,MaekawaEd2012}. It can be generated from non-equilibrium spin states, i.e., spin accumulation and spin dynamics.
The former is routinely produced in nonlocal spin valves\cite{Jedema2001}.
In ferromagnets, the latter is excited by ferromagnetic resonance\cite{Saitoh2006}, temperature gradient\cite{Uchida2008}, and sound waves in the magnetic insulator\cite{UchidaAcoustic}.
Alternatively, spin currents in nonmagnets have been generated by the spin Hall effect\cite{SHE}, in which a strong spin-orbit interaction (SOI) is utilized. 
All these existing methods rely on exchange coupling of spins with local magnetization or on SOI.

In this Letter, we pursue a new route for generating spin currents by considering spin-rotation coupling\cite{SRC}: 
\begin{eqnarray}
H_{S} = - \frac{\hbar}{2} \vsigma \cdot \vOmega,
\end{eqnarray}
where $\hbar \vsigma/2$ is the electron spin angular momentum and $\vOmega$ is the mechanical rotation frequency.
The method requires neither magnetic moments nor SOI. 
In this sense, the mechanism proposed here is particularly relevant in nonmagnets with longer spin lifetime.
%%%%%%
\begin{figure}[htbp]
\begin{center}
\includegraphics[scale=0.6]{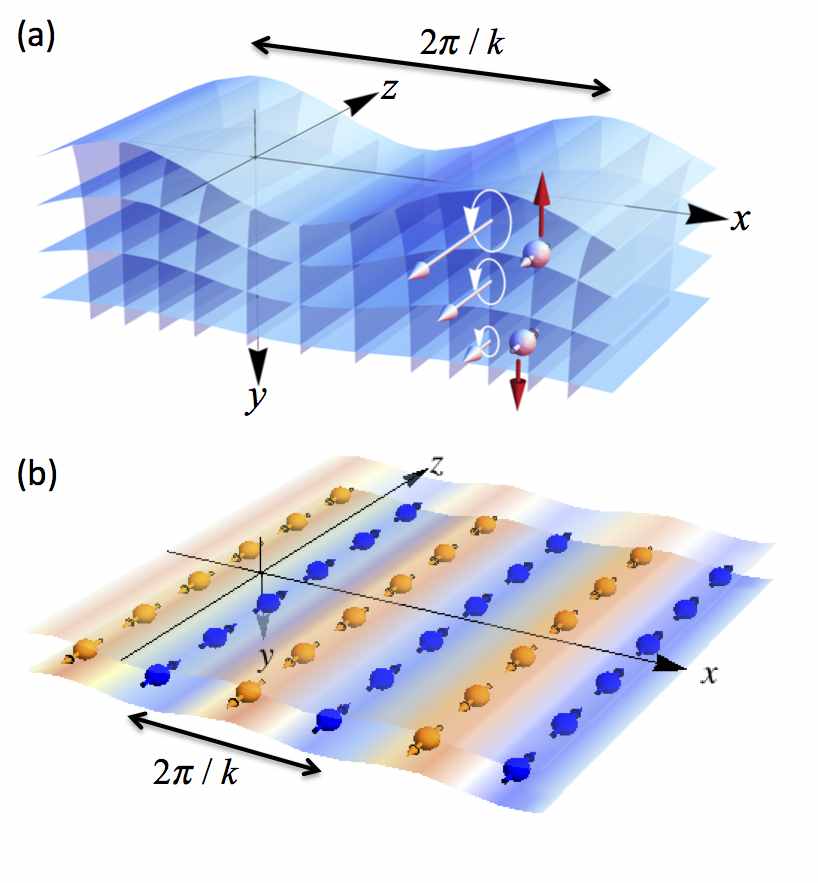}
\end{center}
\caption{Snapshot of mechanical generation of spin current induced by SAW. (a) In presence of a SAW propagating in the $x$-direction, a gradient of mechanical rotation around the $z$-axis is induced. 
The rotation couples to electron spins, and then the $z$-polarized spin current flows in the $y$-direction.  
(b) Spin accumulation induced on the surface. Because spins are polarized parallel to the rotation axis ($\pm z$), the striped pattern of spin accumulation arises at the surface.}
\label{Fig1}
\end{figure}
%%%%%%

\paragraph{Nonuniform rotational motion---.}
Here, we consider rotational motion of the lattice:
\begin{eqnarray}
\vOmega = \frac12 \nabla \times \dot{\v{u}}, \label{OmegaLattice}
\end{eqnarray}
where $\v{u}$ is the displacement vector of the lattice\cite{LandauElasticity}.
When the lattice vibration has transverse modes, Eq. (\ref{OmegaLattice}) does not vanish.
In such a case, the mechanical angular momentum of the lattice can be converted into spin angular momentum via $H_{S}$\cite{SpinTransversePhonon}.
However, as shown later, that $\vOmega$ is finite is insufficient to generate spin currents elastically.
Both the time derivative and the gradient of rotational modes are necessary for the generation of spin current.
For this purpose, we focus on surface acoustic waves (SAWs), which induce rotational deformations that vary in space and time (Fig. \ref{Fig1}).

In presence of SAW, a gradient of mechanical rotation is induced in the attenuation direction. 
We extend the spin diffusion equation to include the coupling between elastic rotation and spin.
By solving the equation in presence of SAW, 
we can evaluate the induced spin current for metals and semiconductors.
%It is also found that the spin current modulates the diffusion constant in the elastic equation of the lattice because of the angular momentum transfer from the lattice into the electron system.

\paragraph{Spin diffusion equation with spin-rotation coupling.---}
First of all,  we examine effects of spin-rotation coupling on spin density. 
When the mechanical rotation, $\Omega$, whose axis is in the $z$-direction, is applied, the electron spins align parallel to the axis of rotation. 
This is known as the Barnett effect\cite{Barnett1915}. 
In this case, the bottom of the energy band of the electron is shifted by $\hbar \Omega/2$.
The number density of up(down) spin electrons is then given by
\begin{eqnarray}
n_{\uparrow (\downarrow)}=\int_{\pm\hbar \Omega/2}^{\mu_{\uparrow (\downarrow)}} d\varepsilon N_{0}(\varepsilon),
\end{eqnarray}
where $N_{0}$ is the density of states for electrons, and $\mu_{\uparrow}$ and $\mu_{\downarrow}$ are chemical potentials for up and down spin electrons, respectively. The $z$-direction is selected as the quantization axis.
Then, spin density can be estimated as
\begin{eqnarray}
n_{\uparrow} - n_{\downarrow} \approx N_{0} (\delta \mu - \hbar \Omega), 
\end{eqnarray}
where $\delta \mu = \mu_{\uparrow} - \mu_{\downarrow}$ is spin accumulation.
Here, a constant density of state is assumed for simplicity.
Spin relaxation occurs in two processes: one is on-site spin flip with the spin lifetime, $\tau_{\rm sf}$, and the other is spin diffusion with the diffusion constant, $D$. 
Equating these processes leads to 
$\del_{t} (n_{\uparrow}-n_{\downarrow}) = \tau_{\rm sf}^{-1}N_{0} \delta \mu   + D \nabla^{2}(N_{0}\delta \mu)$.
Next, we obtain the extended spin diffusion equation in presence of spin-rotation coupling:
\begin{eqnarray}
\left(\del_{t}  - D\nabla^{2} + \tau_{\rm sf}^{-1}  \right) \delta \mu = \hbar \del_{t}\Omega,  \label{ESDeq}
\end{eqnarray}
The R.H.S. of Eq. (\ref{ESDeq}) is a source term originating from spin-rotation coupling. 
$Z$-polarized spin current can be calculated from the solution of Eq. (\ref{ESDeq}) as
\begin{eqnarray}
\v{J}_{s}^{z} = \frac{\sigma_{0}}{e} \nabla \delta \mu,\label{SpinCurrent}
\end{eqnarray}
with conductivity $\sigma_{0}$.
If the mechanical rotation is constant with time, the source term vanishes. 
Moreover, even if mechanical rotation depends on time, the uniform rotation in space cannot generate spin currents 
because spin accumulation is independent of space.

\paragraph{Spin accumulation induced by SAW.---}
Let us consider generation of spin current due to spin-rotation coupling of SAWs in nonmagnetic metals or semiconductors.
Our setup is shown in Fig. \ref{Fig1} (a). SAWs are generated in the $xz$-plane and penetrates a nonmagnetic material along the $y$-direction.
They then induce mechanical rotation around the $z$-axis, whose frequency $\vOmega=(0,0,\Omega)$ is given by\cite{LandauElasticity}
\begin{eqnarray}
\Omega(x,y,t) =\frac{\omega^{2} u_{0}}{2c_{t}} \exp \{ -k_{t}y + i(kx- \omega t)\},\label{OmegaSAW}
\end{eqnarray}
where $\omega$ and $u_{0}$ are the frequency and amplitude of the mechanical resonator, $k$ is wave number, $c_{t}$ is the transverse sound velocity, and $k_{t}$ is the transverse wave number. 
The frequency $\omega$ is related to the wave number as $\omega = c_{t} k \xi$ and the transverse wave number as $k_{t} = k\sqrt{1-\xi^{2}}$, where $\xi$ satisfies the equation $\xi^{6} - 8\xi^{4} + 8\xi^{3}(3-2c_{t}^{2}/c_{l}^{2})-16(1-c_{t}^{2}/c_{l}^{2})=0$ and $c_{l}$ is the longitudinal sound velocity. 
The Poisson ratio, $\nu$, is related to the ratio of velocities as $(c_{t}/c_{l})^{2}=(1-2\nu)/2(1-\nu)$, and $\nu$ and $\xi$ are related as $\xi \approx (0.875+ 1.12\nu)/(1+\nu)$.

Spin accumulation generated by the SAW can be evaluated by solving Eq. (\ref{ESDeq}). 
By inserting Eq. (\ref{OmegaSAW}) and $\delta \mu = \delta \mu_{y}(y,t)e^{ik x}$ into Eq. (\ref{ESDeq}), the spin diffusion equation can be rewritten as
\begin{eqnarray}
\left( \del_{t} - D \del^{2}_{y} + \ttsf^{-1}  \right) \delta \mu_{y} (y,t) = - i \omega \hbar\Omega_{0} e^{-k_{t}y - i \omega t},\label{ESDeq2}
\end{eqnarray}
where $\ttsf =\tsf(1+\ls^{2} k^{2})^{-1}$ with the spin diffusion length $\ls=\sqrt{D\tsf}$ and $\Omega_{0}=\omega^{2}u_{0}/2c_{t}$.
With the boundary condition $\del_{y} \delta \mu = 0$ on the surface $y=0$, 
the solution is given by
\begin{eqnarray}
\delta \mu_{y}(y,t) =&& - i \omega \hbar\Omega_{0}
 \int_{0}^{\infty}dt' \int_{0}^{\infty}dy' \frac{ \theta(t-t') e^{-(t-t')/\ttsf}}{\sqrt{4\pi D(t-t')}} \nonumber\\
&&\ \times 
(e^{-\frac{(y-y')^{2}}{4D(t-t')}}+e^{-\frac{(y+y')^{2}}{4D(t-t')}})  e^{-k_{t}y' - i \omega t'}. \label{SolAccumulation}
\end{eqnarray}

%%%%%%
Here, let us consider the time evolution of spin accumulation at the surface, $y=0$.   
Because each spin aligns parallel to the rotation axis, i.e., the $\pm z$-axis, 
a striped pattern of spin accumulation [shown in Fig. \ref{Fig1} (b)] arises at the surface. 
The period of spatial pattern is the same as the wavelength of SAW, $2\pi/k$. 

Recently, spin precession controlled by SAW was observed by using the time-resolved polar megneto-optic Kerr effect (MOKE)\cite{Hernandez2010,Sanada2011}.
In our case, in-plane spin polarization is induced. 
Therefore, transversal or longitudinal MOKE can be used to observe patterns shown in Fig. \ref{Fig1} (b).

\begin{figure}[htbp]
\begin{center}
\includegraphics[scale=0.6]{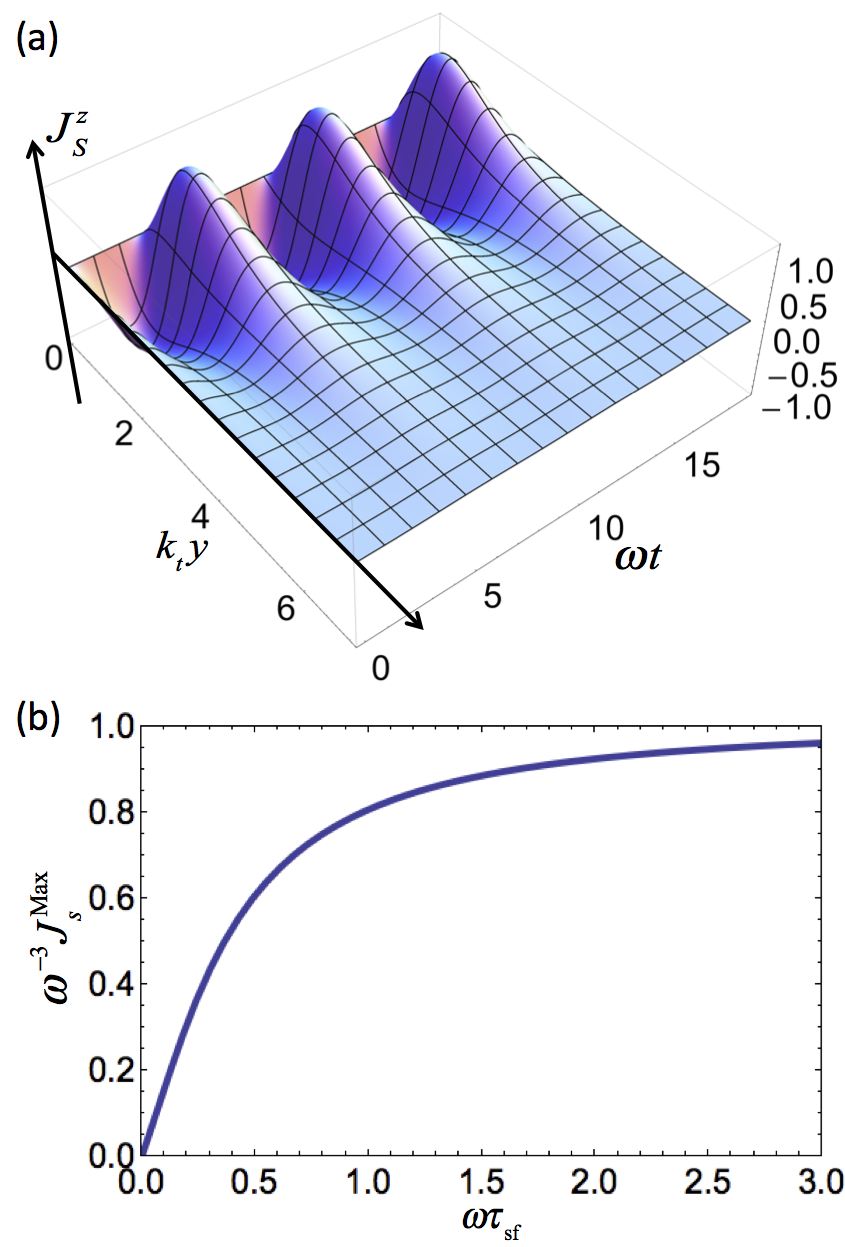} %\includegraphics[scale=0.5]{deltamu.eps}
\end{center}
\caption{(a) Spin current induced by SAW $J_{s}^{z}$ plotted as a function of $ k_{t}y$ and $\omega t$ for fixed $x$ and $z$. The spin current oscillates with time. Maximum amplitude is located near the surface, $k_{t}y \approx 1$. (b) Maximum amplitude of the spin current scaled by $\omega^{3}$ plotted as a function of $\omega \tau_{\rm sf}$. When $\omega \tsf  \ll 1$, the scaled amplitude, $J_{s}^{\rm Max} \omega^{-3}$, increases linearly. On the other hand, it saturates when $\omega \tsf \gg 1$. Accordingly, the maximum amplitude, $J_{s}^{\rm Max}$, is proportional to $\omega^{4}$ in the former case, whereas $J_{s}^{\rm Max} \propto \omega^{3}$ in the latter case.}
\label{Fig2}
\end{figure}

%%%%%%
\paragraph{Spin current from SAW.---}
From Eqs. (\ref{SpinCurrent}) and (\ref{SolAccumulation}) we obtain $z$-polarized spin current in the $y$-direction. 
In Fig. \ref{Fig2}, the SAW-induced spin current is shown. 
The spin current, $J_{s}^{z}$, is plotted as a function of $ \kt y$ and $\omega t$ in Fig. \ref{Fig2} (a). The spin current oscillates with the same frequency as that of the mechanical resonator, $\omega$. The maximum amplitude, $J_{s}^{\rm Max}$, is found near the surface, $k_{t} y \approx 1$. 
In Fig. \ref{Fig2} (b), maximum amplitude scaled by $\omega^{3}$ is plotted as a function of $\omega \tsf$. The amplitude increases linearly when $\omega \tsf \ll 1$, whereas it saturates when $\tsf \gg \omega^{-1}$. 
In other words, $J_{s}^{\rm Max} \propto \omega^{4}$ for $\omega \tsf \ll 1$ whereas $J_{s}^{\rm Max} \propto \omega^{3}$ for $\omega \tsf \gg1$.   
%%%%%%

To clarify material dependence of spin current, we use an asymptotic solution of Eq. (\ref{ESDeq}) for $\kt y \gg 1$:
\begin{eqnarray}
\delta \mu \approx \frac{i\omega \ttsf}{i\omega \ttsf+\ls^{2} k_{t}^{2} -1}\hbar \Omega_{0} e^{-k_{t}y + i(kx-\omega t)}, \label{Asym}
\end{eqnarray}
which leads to 
\begin{eqnarray}
J_{s}^{z} \approx \frac{-i\omega \ttsf}{i\omega \ttsf+\ls^{2} k_{t}^{2} -1} \frac{\hbar \sigma_{0}}{2e} \frac{\omega^{3} u_{0}}{c_{t}^{2}}\frac{\sqrt{1-\xi^{2}}}{\xi} e^{-k_{t}y + i(kx-\omega t)} \nonumber \\
\end{eqnarray}
where $\Omega_{0} = \omega^{2}u_{0}/2c_{t}$. 

When spin relaxation is absent, $\omega \ttsf \gg 1$, one obtains 
\begin{eqnarray}
\delta \mu \approx \hbar \Omega_{0} e^{-k_{t}y + i(kx-\omega t)}
\end{eqnarray}
and 
\begin{eqnarray}
J_{s}^{z} \approx -(\sigma_{0}/e)k_{t} \hbar \Omega_{0} e^{-k_{t}y + i(kx-\omega t)}.
\end{eqnarray}

When $\omega \ttsf \ll 1$ and $\ls \kt \ll1$, the spin current becomes
\begin{eqnarray}
J_{s}^{z} \approx \omega \tau_{\rm sf}\cdot \frac{\hbar \sigma_{0}}{2e} \frac{\omega^{3} u_{0}}{c_{t}^{2}}\frac{\sqrt{1-\xi^{2}}}{\xi}e^{-k_{t}y + i(kx- \omega t + \pi/2 )}.\label{AsymJs}
%J_{s}^{z} \approx   \frac{\hbar \sigma_{0}}{2e} \frac{\omega^{3} u_{0}}{c_{t}^{2}}\frac{1-\xi^{2}}{\xi}\exp\left\{-k_{t}y + i\left(kx- \omega t + \frac{\pi}2 \right)\right\}
\end{eqnarray}
%Since the poisson's ratios of metals and semiconductors including are almost same, 
%the amplitude of the spin current 
%\begin{eqnarray}
%\frac{J_{s}^{z}}{J_{s}^{z,{\rm Pt}}} \approx \frac{\tsf}{\tsf^{\rm Pt}}\cdot \frac{\sigma_{0}}{\sigma_{0}^{\rm Pt}} \cdot \left(  \frac{c_{t}}{c_{t}^{\rm Pt}}\right)^{-2}.\label{AsymRatio}
%\end{eqnarray}
As seen in Eq. (\ref{AsymJs}), the larger spin current can be obtained from materials with the longer spin lifetime, namely, weaker SOI.

Let us examine the SAW-induced spin current in typical nonmagnetic materials. 
Using Eqs. (\ref{SpinCurrent}) and (\ref{SolAccumulation}), 
the maximum value of the spin current for Al, Cu, Ag, Au, and n-doped GaAs normalized by that of Pt, $J_{s}^{\rm  Max,Pt}$,
is computed as listed in Table 1. The ratio of the maximum amplitude of the spin current to that of Pt, $\bar{J}_{s}$, is defined as $\bar{J}_{s}=J_{s}^{\rm Max}/J_{s}^{\rm Max,Pt}$. The ratio depends mainly on the conductivity, $\sigma_{0}$, and spin lifetime, $\tsf$. 
The order, $\bar J_{s}^{\rm Cu} > \bar J_{s}^{\rm Al} > \bar J_{s}^{\rm Ag} > \bar J_{s}^{\rm Au} > \bar J_{s}^{\rm Pt}=1$, is unchanged, since these materials well satisfy $\omega \tsf \ll 1$. 
For GaAs, the ratio, $\bar J_{s}^{\rm GaAs}$, is greater than 1 for $\omega <$1GHz, whereas it becomes smaller than 1 for $\omega >$1GHz. 
This happens because the spin lifetime of GaAs is much longer than that of Pt; i,e.,  
the dimensionless parameter, $\omega \tsf$, becomes much greater than 1 when $\omega > 1$GHz. 
In such a case, $J_{s}^{\rm Max}/\omega^{3}$ for GaAs saturates, whereas that for Pt linearly increases, as shown in Fig. \ref{Fig2} (b). 
  
It is worth noting that the spin current generated in a metal with weak spin-orbit interaction such as those of Al and Cu is much larger than that in Pt. 
In addition, the spin current in n-doped GaAs is comparable to that in Pt. 
Although conductivities of semiconductors are much smaller than those of metals, the spin lifetime is much longer.
Hence, the amplitude of the induced spin current in GaAs is comparable to that in Pt for $\omega \tsf^{\rm GaAs} <1$. 

Recently, SAWs in the GHz frequency range have been used for spin manipulation\cite{Sanada2011,Weiler}.
Here, we evaluate spin current at such high frequencies. 
In case of $u_{0}=10^{-9}$m, $\omega/2\pi = 10$GHz,
Pt has the maximum amplitude, $J_{s}^{z,\rm Pt}\approx 4 \times 10^{6}$A/m$^{2}$.

Conventionally, generation of spin current in nonmagnetic materials has required strong SOI 
because the spin Hall effect has been utilized. In other words, nonmagnetic materials with short spin lifetimes have been used. 
On the contrary, the mechanism proposed here requires longer spin lifetimes to generate larger spin currents. 
Therefore, more options are available for spin-current generation in nonmagnets than ever before.

%%%%%%
\paragraph{Enhancement of the SAW-induced spin current.---}
Very recently, it has been predicted that spin-rotation coupling can be enhanced by an interband mixing of solids\cite{MamoruSRC}:
\begin{eqnarray}
H_{S}' =  -(1+\delta g)\frac{\hbar}{2}\vsigma \cdot \vOmega.
\end{eqnarray}
Here, $\delta g$ is given by $\delta g = g-g_{0}$ where $g_{0}=2$ and $g$ are electron $g$ factors in vacuum and solids, respectively. 
Considering enhancement, 
the mechanical rotation, $\Omega$, inserted into the extended spin diffusion equation, Eq. (\ref{ESDeq}), is replaced by $(1+\delta g)\Omega$.
Consequently, 
the spin accumulation, $\delta \mu$, is modified as
$\delta \mu \to (1+ \delta g)\delta \mu$, and accordingly, 
the induced spin current as $J_{s}^{z} \to (1+\delta g)J_{s}^{z}$.
For lightly doped n-InSb at low temperature, $g \approx -49$ has been employed in a recent experiment\cite{Jaworski2012}. 
In this case, one obtains $\delta g \approx -51$. Therefore, the amplitude of the spin current can be 50 times larger.

\begin{widetext}
\begin{center}
\begin{table}[b]
\caption{SAW-induced spin current for Pt, Al, Cu, Ag, Au, and GaAs. The ratio is given by $\bar{J}_{s} = J_{s}^{\rm Max}/J_{s}^{\rm Max, Pt}$, where $J_{s}^{\rm Max,Pt}$ is the maximum amplitude of the spin current for Pt. 
The ratio $\bar{J}_{s}$ depends on the Poisson's ratio, $\nu$, the transverse velocity, $c_{t}$, conductivity, $\sigma_{0}$, and spin lifetime, $\tsf$. }
\begin{tabular}{|c||c|c|c|c||c|c|c|c||c|}\hline
	&	$\nu $ &	$c_{t}$[m/s] &	$\sigma_{0}$[$10^{7}(\Omega {\rm m})^{-1}$] & $\tsf$ [ps] & $\bar{J}_{s}$(0.1GHz)  	& $\bar{J}_{s}$(1GHz) & $\bar{J}_{s}$(2.5GHz)	& $\bar{J}_{s}$(10GHz)  & Ref. %&	$\delta \mu/\delta \mu^{\rm Pt}$
	 \\ \hline
Pt	&	0.377&	1730&	0.96 & 0.3 	&			1  &	1 & 1	& 1  	& \cite{Vila2007}	\\
Al	&	0.345 &	3040	&	1.7 &	    100 &			390	& 290  & 210 &  62	&  \cite{Jedema2002}	 \\	
Cu	&	0.343&	2270&	7.0 &     42	&		950	& 700 & 650 & 330	 & \cite{Jedema2001}	 \\
Ag	&	0.367&	1660&	2.9 &	  3.5&				44 	& 38	& 34	& 32 & \cite{Godfrey2006} \\
Au	&	0.44 &	1220&	2.5 &  2.8	&			 	42	& 35	& 33 & 30 &\cite{Ku2006}	 \\
GaAs&	0.31  &	2486& 3.3$\times 10^{-4}$ & $10^{5}$&	1.6	& 0.13	& 0.050 & 0.013	 & \cite{Kikkawa1998}	  \\

\hline
\end{tabular}
\end{table}
%%%%%%
\end{center}
\end{widetext}

%%%%%%
\paragraph{Discussion and conclusion.---}
The method of spin-current generation using spin-rotation coupling is purely of mechanical origin; i.e., it is independent of exchange coupling and SOI. 
Lattice dynamics directly excites the nonequilibrium state of electron spins, and consequently, 
spin current can be generated in nonmagnets. 

As an example, we have theoretically demonstrated that SAW, a situation in which rotational motion of lattice couples with electron spins, can be exploited for spin current generation. 
The spin diffusion equation is extended to include effects of spin-rotation coupling.
The solution of the equation reveals that the spin current is generated parallel to the gradient of the rotation.
Moreover, it has been determined that larger spin current can be generated in nonmagnetic materials with longer spin lifetimes.
This means that Al and Cu, which have been considered as good materials for a spin conducting channel, 
are favorable for generating spin current. 
Spin accumulation induced by the SAW on the surface will be observed by Kerr spectroscopy. 

These results can be generalized for other lattice dynamics. 
SAW discussed above is the Rayleigh wave, which induces rotation with the axis parallel to the surface. 
For instance, the Love wave\cite{Love1967}, horizontally polarized shear wave, can be utilized to generate spin currents whose spin polarization is perpendicular to the surface. 
The use of spin rotation coupling, argued here, opens up a new pathway for creating spin currents by elastic waves.

\begin{acknowledgments}
The authors thank S. Takahashi for valuable discussions.
% Ito-san, Muon people
This study was supported by a Grant-in-Aid for Scientific Research from MEXT.
\end{acknowledgments}


\begin{thebibliography}{10}

%\bibitem{Zutic2004}I. Zutic, J. Fabian, and S. Das Sarma, Rev. Mod. Phys. {\bf 76}, 323 (2004).
\bibitem{MaekawaEd2006}S. Maekawa, ed., {\it Concepts in Spin Electronics} (Oxford University Press, Oxford, 2006).
\bibitem{MaekawaEd2012}S. Maekawa, S. Valenzuela, E. Saitoh, and T. Kimura, ed., {\it Spin Current} (Oxford University Press, Oxford, 2012).

\bibitem{Jedema2001}F. J. Jedema, A. T. Filip, and B. J. van Wees, Nature (London), {\bf 410}, 345 (2001).
%\bibitem{Hirohata2001}A. Hirohata, Y. B. Xu, C. M. Guertler, and J. A. C. Bland, Phys. Rev.  B{\bf 62}, 104425 (2001).
%\bibitem{Werake2010}L. K. Werake and H. Zhao, Nat. Phys. {\bf 6}, 875 (2010).




%\bibitem{Ando2008}K. Ando, S. Takahashi, K. Harii, K. Sasage, J. Ieda, S. Maekawa, and E. Saitoh, Phys. Rev. Lett. {\bf 101}, 036601 (2008).
%\bibitem{Barnes2007}S. E. Barnes and S. Maekawa, Phys. Rev. Lett. {\bf 98}, 246601 (2007).

%\bibitem{Brataas2002}A. Brataas, Y. Tserkovnyak, G. E. W. Bauer, and B. I. Halperin, Phys. Rev. {\bf B 66}, 060404(R) (2002).
\bibitem{Saitoh2006}E. Saitoh, M. Ueda, H. Miyajima, and G. Tatara. Appl. Phys. Lett. {\bf 88}, 182509 (2006).
%\bibitem{SHE}
%M. I. Dyakonov and V. I. Perel, Phys. Lett. A35, 459 (1971);
%J. E. Hirsch, Phys. Rev. Lett. 83, 1834 (1999);
%S. Zhang, Phys. Rev. Lett. 85, 393  (2000).
%S. Murakami, N. Nagaosa, and S. C. Zhang, Science 301, 1348  (2003);
%J. Sinova, D. Culcer, Q. Niu, N. A. Sinitsyn, T. Jungwirth, and A. H.
%MacDonald, Phys. Rev. Lett. 92, 126603  (2004).
%E. I. Rashba, Phys. Rev. B 68, 241315 (2003).
%J. Schliemann and D. Loss, Phys. Rev. B 69, 165315  (2004).
%E. G. Mishchenko, A. V. Shytov, and B. I. Halperin, Phys. Rev. Lett. 93, 226602 (2004).
%K. Nomura, J. Sinova, T. Jungwirth, Q. Niu, and A. H. MacDonald, Phys. Rev. B 71, 041304 (2005).
%T. Kimura, Y. Otani, K. Tsukagoshi, and Y. Aoyagi, J. Magn. Magn. Mater. 272?276, E1333  (2004).


\bibitem{Uchida2008}K. Uchida, S. Takahashi, K. Harii, J. Ieda, W. Koshibae,
K. Ando, S. Maekawa, and E. Saitoh, Nature (London) {\bf 455}, 778
(2008).
\bibitem{UchidaAcoustic}
K. Uchida, H. Adachi, T. An, T. Ota, M. Toda, B. Hillebrands, S. Maekawa, and E. Saitoh, Nature Mater. {\bf 10}, 737 (2011); K. Uchida, T. An, Y. Kajiwara , M. Toda , and E. Saitoh, Appl. Phys. Lett.  {\bf 99}, 212501 (2011); K. Uchida, H. Adachi, T. An, H. Nakayama, M. Toda, B. Hillebrands, S. Maekawa, and E. Saitoh, J. Appl. Phys. {\bf 111}, 053903 (2012).
%\bibitem{Uchida2011nm}K. Uchida, H. Adachi, T. An, T. Ota, M. Toda, B. Hillebrands, S. Maekawa, and E. Saitoh, Nature Mater. {\bf 10}, 737 (2011).
%\bibitem{Uchida2011apl}K. Uchida, T. An, Y. Kajiwara , M. Toda , and E. Saitoh, Appl. Phys. Lett.  {\bf 99}, 212501 (2011). 
%\bibitem{Uchida2012jap}K. Uchida, H. Adachi, T. An, H. Nakayama, M. Toda, B. Hillebrands, S. Maekawa, and E. Saitoh, J. Appl. Phys. {\bf 111}, 053903 (2012).


\bibitem{SHE}Y. K. Kato, R. C. Myers, A. C. Gossard, and D. D. Awschalom, Science {\bf 306}, 1910 (2004); J. Wunderlich, B. Kaestner, J. Sinova, and T. Jungwirth, Phys. Rev. Lett. {\bf 94}, 047204 (2005); T. Kimura, Y. Otani, T. Sato, S. Takahashi, and S. Maekawa, Phys. Rev. Lett. {\bf 98}, 156601 (2007).



%\bibitem{Weiler2011}M. Weiler, L. Dreher, C. Heeg, H. Huebl, R. Gross, M. S. Brandt, and S. T. B. Goennenwein, Phys. Rev. Lett. {\bf 106}, 117601 (2011).
%\bibitem{Weiler2012}M. Weiler, H. Huebl, F. S. Goerg, F. D. Czeschka, R. Gross, and S. T. B. Goennenwein, Phys. Rev. Lett. {\bf 108}, 176601 (2012). 
%\bibitem{Maekawa1976}S. Maekawa and M. Tachiki, AIP Conf. Proc. 29, 542 (1976).
%\bibitem{Ganguly1976}A. K. Ganguly, K. L. Davis, D. C. Webb, and C. Vittoria, J. Appl. Phys. {\bf 47}, 2696 (1976).





%\bibitem{Rugar1992}D. Rugar, C. S. Yannoni, and J. A. Sidles, Nature (London) {\bf 360}, 563 (1992).
%\bibitem{Wallis2006}T. M. Wallis, J. Moreland, and P. Kabos, Appl. Phys. Lett. {\bf 89}, 122502 (2006).
%\bibitem{Zolfagharkhani2008}G. Zolfagharkhani \emph{et al.}, Nat. Nanotechnol. {\bf 3}, 720 (2008).

%\bibitem{Tejada2010}J. Tejada, R. D. Zysler, E. Molins, and E. M. Chudnovsky, Phys. Rev. Lett. {\bf 104}, 027202 (2010).%Evidence for Quantization of Mechanical Rotation of Magnetic Nanoparticles




%\bibitem{Mamoru2011}M. Matsuo, J. Ieda, E. Saitoh, and S. Maekawa, Phys. Rev. Lett. {\bf 106}, 076601 (2011).
%\bibitem{Mamoru2011a}M. Matsuo, J. Ieda, E. Saitoh, and S. Maekawa, Appl. Phys. Lett. {\bf 98}, 242501 (2011).
%\bibitem{Mamoru2011b}M. Matsuo, J. Ieda, E. Saitoh, and S. Maekawa, Phys. Rev. {\bf B 84}, 104410 (2011).

%Spin-Rotation coupling

\bibitem{SRC}C. G. de Oliveira and J. Tiomno, Nuovo Cimento {\bf 24}, 672 (1962); B. Mashhoon, Phys. Rev. Lett. {\bf 61}, 2639 (1988); J. Anandan, Phys. Rev. Lett. {\bf 68}, 3809 (1992); B. Mashhoon, Phys. Rev. Lett. {\bf 68}, 3812 (1992); F. W. Hehl and W.-T. Ni, Phys. Rev. {\bf D42}, 2045 (1990).

%\bibitem{Oliveira1962}C. G. de Oliveira and J. Tiomno, Nuovo Cimento {\bf 24}, 672 (1962).
%\bibitem{Mashhoon1988} B. Mashhoon, Phys. Rev. Lett. {\bf 61}, 2639 (1988); see also J. Anandan, Phys. Rev. Lett. {\bf 68}, 3809 (1992); B. Mashhoon, Phys. Rev. Lett. {\bf 68}, 3812 (1992).
%\bibitem{Hehl1990} F. W. Hehl and W.-T. Ni, Phys. Rev. {\bf D42}, 2045 (1990).

%\bibitem{Einstein-deHaas1915}A. Einstein and W. J. de Haas, Verh. Dtsch. Phys. Ges. {\bf 17}, 152 (1915).
%\bibitem{Barnett1935}S. J. Barnett, Rev. Mod. Phys. {\bf 7}, (1935) 129.

\bibitem{LandauElasticity}L. D. Landau and E. M. Lifshitz, {\it Theory of Elasticity}
(Pergamon, New York, 1959).

\bibitem{SpinTransversePhonon}
E. M. Chudnovsky, D. A. Garanin, and R. Schilling, Phys. Rev. {\bf B 72}, 094426 (2005); %Universal mechanism of spin relaxation in solids, rigid spin cluster in an elastic medium in presence of the magnetic field.
C. Calero, E. M. Chudnovsky, and D. A. Garanin, Phys. Rev. Lett. {\bf 95}, 166603 (2005); %Quantum dot
C. Calero and E. M. Chudnovsky, Phys. Rev. Lett. {\bf 99}, 047201 (2007). % molecular magnet and Rabi oscillation 
%\bibitem{Calero2005prl}C. Calero, E. M. Chudnovsky, and D. A. Garanin, Phys. Rev. Lett. {\bf 95}, 166603 (2005). %Quantum dot
%\bibitem{Calero2007prl}C. Calero and E. M. Chudnovsky, Phys. Rev. Lett. {\bf 99}, 047201 (2007). % molecular magnet and Rabi oscillation 



\bibitem{Barnett1915}S. J. Barnett, Phys. Rev. {\bf 6}, (1915) 239.


\bibitem{Hernandez2010}A. Hern\'andez-M\'inguez, K. Biermann, S. Lazi\'c, R. Hey, and P. V. Santos, Appl. Phys. Lett. {\bf 97}, 242110 (2010).
\bibitem{Sanada2011}H. Sanada, T. Sogawa, H. Gotoh, K. Onomitsu, M. Kohda, J. Nitta, and P. V. Santos, Phys. Rev. Lett. {\bf 106}, 216602 (2011).

\bibitem{Weiler}M. Weiler, L. Dreher, C. Heeg, H. Huebl, R. Gross, M. S. Brandt, and S.
T. B. Goennenwein, Phys. Rev. Lett. {\bf 106}, 117601 (2011).
%; M. Weiler, H. Huebl, F. S. Goerg, F. D. Czeschka, R. Gross, and S. T. B. Goennenwein, Phys. Rev. Lett. {\bf 108}, 176601 (2012). 

%Refs on spin lifetime
\bibitem{Vila2007}L. Vila, T. Kimura, and Y. C. Otani, Phys. Rev. Lett. {\bf 99}, 226604 (2007).


\bibitem{Jedema2002}F. J. Jedema, H. B. Heersche, A. T. Filip, J. J. A. Baselmans, and B. J. van Wees, Nature (London), {\bf 416}, 713 (2002).% Al in nonlocal spin valve

\bibitem{Godfrey2006}R. Godfrey and M. Johnson, Phys. Rev. Lett. {\bf 96}, 136601 (2006).
\bibitem{Ku2006}J.-H. Ku, J. Chang, H. Kim, and J. Eom, Appl. Phys. Lett. {\bf 88}, 172510 (2006).
\bibitem{Kikkawa1998}J. M. Kikkawa and D. D. Awschalom, Phys. Rev. Lett. {\bf 80}, 4313 (1998).



\bibitem{MamoruSRC}M. Matsuo, J. Ieda, and S. Maekawa, arXiv:1211.0127.
%\bibitem{Kane1957}E. O. Kane, J. Phys. Chem. Solids {\bf 1}, 249 (1957).
\bibitem{Jaworski2012}C. M. Jaworski, R. C. Myers, E. Johnston-Halperin, and J. P. Heremans, Nature (London) {\bf 484}, 210 (2012).


%\bibitem{Tserkovnyak2002}Y. Tserkovnyak, A. Brataas, and G. E. W. Bauer, Phys. Rev. Lett. {\bf 88}, 117601 (2002). %spin pumping

\bibitem{Love1967}A. E. H. Love, {\it A Treatise on the Mathematical Theory of Elasticity} (Dover, New York, USA, 1967). 
%(Chapter 11: Theory of the propagation of seismic waves)




\end{thebibliography}
\end{document}